\documentclass[11pt,a4paper]{article}
\usepackage{amsmath}
\usepackage{epsfig}
\def\a{\alpha}
\def\b{\beta}
\def\c{\gamma}
\def\d{\theta}
\author{H. Mohseni Sadjadi\footnote{mohseni@phymail.ut.ac.ir} and M.
Alimohammadi\footnote{alimohmd@ut.ac.ir}\\
{\small Department of Physics, University of Tehran,}
\\ {\small North Karegar Ave. Tehran, Iran.}}

\title{Cosmological coincidence problem in interacting dark energy models }
\date{}
\begin{document}
\maketitle
\begin{abstract}
The interacting dark energy model with interaction term $Q=
\lambda_m H\rho_m+\lambda_dH\rho_d$ is considered. By studying the
model near the transition time, in which the system crosses the
$\omega=-1$ phantom-divide-line, the conditions needed to overcome
the coincidence problem is investigated. The phantom model, as a
candidate for dark energy, is considered, and for two specific
examples, the quadratic and exponential phantom potentials, it is
shown that it is possible the system crosses the $\omega=-1$ line,
meanwhile the coincidence problem is alleviated, the two facts
that have root in observations.
\end{abstract}

\section{Introduction}
Nowadays based on astrophysical data it is believed that the
universe is accelerating \cite{acc}. The origin of this
acceleration is still unknown and different models have been
proposed to elucidate this subject. One picture is the assumption
that nearly 70\% of the universe is composed of a smooth energy
component with negative pressure dubbed as dark energy.  A simple
candidate for dark energy is the cosmological constant \cite{cc1}
which suffers from conceptual problems such as fine-tuning and
coincidence problems \cite{cc2}. Therefore alternative models,
e.g., introducing dynamical exotic fields such as scalar fields
with suitably chosen potentials, have been introduced
\cite{field}.

In dark energy models, the ratio of  matter to dark energy
density, $r$, is expected to decrease rapidly (proportional to the
scale factor) as the universe expands, but observations show that
these densities are of the same order today. To solve this problem
(known as coincidence problem), one can adopt an evolving dark
energy field with suitable non-gravitational interaction with
matter \cite{inter},\cite{LP}. Various models corresponding to
different forms of interaction, leading to a constant or slowly
varying (soft coincidence) $r$ at late times, have been proposed
\cite{r}.

Some present data seems to favor an evolving dark energy,
corresponding to an equation of state (EOS) parameter less than
$\omega = -1$ at present epoch (phantom regime)  from $\omega>-1$
in the near past (quintessence regime) \cite{cross}. So another
cosmological coincidence problem may be proposed: why $\omega= -1$
crossing is occurred at the present time \cite{seccoin}.

In \cite{vik}, it was shown that  $\omega= -1$ crossing in models
including matter and phantom scalar field is either impossible or
unstable with respect to cosmological perturbations. However, this
transition may be possible for scalar-tensor theories \cite{st},
multi-field models \cite{mf}, and coupled dark energy models with
specific couplings \cite{coup1},\cite{coup2}.

In \cite{moh}, the transition from quintessence to the phantom
phase in the quintom model was considered in the slow roll
approximation. By studying the Friedman equations near the
transition time, it was shown that in non-interacting quintom
model, $r\simeq 0$ at transition time. This lies in the fact that
the main part of the dark energy at transition time corresponds to
the quintom potential. By considering interaction between cold
dark matter and dark energy, the mutual energy exchange between
two fluids will be allowed and the coincidence problem may be
alleviated.

In this paper we consider dark energy model composed of a phantom
scalar field interacting with cold dark matter. We try to
elucidate the connection between the coincidence problem and
$\omega=-1$ crossing (second cosmological coincidence problem).

It may be worth noting that the phantom models suffer from the
quantum instability problem. Because the phantom fields have
negative kinetic energy, it is possible that a phantom particle
decays into an arbitrary number of phantoms and ordinary
particles, such as gravitons. It can be shown that the decay rates
of these interactions are infinite, which indicates that the
phantom models are dramatically unstable. But if we think of these
models as the low-energy effective theories, with the fundamental
fields having positive kinetic energy, then we should use a
momentum cutoff $\Lambda$ in calculating the decay rates. In this
way, it can be shown that, for $\Lambda \sim M_{\rm pl}$, the
lifetimes can become larger than the age of the universe when one
chooses suitable phantom-gravity interaction potentials, and this
removes the quantum instability of these kinds of phantom models
\cite{quantum}.

The Scheme of the paper is as follows. After the Introduction, we
consider the dark energy model with interaction term $Q= \lambda_m
H\rho_m+\lambda_dH\rho_d$ in section two. By restricting ourselves
to times $t<<h_0^{-1}$ around the transition time ( $h_0$ is the
Hubble parameter), we study the general properties of interacting
dark energy models and the necessary conditions needed to cross
$\omega =-1$ line are obtained. These results are insensitive to
the origin of the dark energy. In section three we assume that
dark energy is composed of phantom scalar field interacting with
cold dark matter. After a general discussion, we illustrate, via
two specific examples, how the necessary conditions for $\omega
=-1$ crossing can alleviate the coincidence problem. It is seen
that it is possible to tune the parameters such that $r_0=3/7$ at
transition time.

We use units $\hbar=c=G=1$ throughout the paper.

\section{$\omega=-1$ crossing in interacting dark energy model}
We consider a spatially flat Friedmann-Lemaitre-Robertson-Walker
(FLRW) universe containing dark energy and dark matter fluids. In
terms of dark energy density $\rho_d$ and matter energy density
$\rho_m$, the Hubble parameter is given by Friedmann equation
\begin{equation}\label{1}
H^2=\frac{8\pi}{3}\rho=\frac{8\pi}{3}(\rho_m+\rho_d),
\end{equation}
where $\rho$ is the total energy density. By introducing
$\Omega_d=\rho_d/\rho$ and $\Omega_m=\rho_m/\rho$, eq.(\ref{1})
can be written as $\Omega_d+\Omega_m=1$, which indicates that the
universe is spatially flat. The derivative of the Hubble parameter
with respect to the comoving time can be extracted from Einstein
equations. The result is
\begin{equation}\label{2}
\dot{H}=-4\pi(\rho_d+P_d+\rho_m).
\end{equation}
$P_d$ is the pressure of the dark energy fluid and the dark matter
is assumed to be pressureless. The equation of state of the
universe is $P=\omega\rho$, where $P=P_d$ is the pressure and
$\omega$ is the equation of state parameter which can be written
as
\begin{equation}\label{3}
\omega=-1-\frac{2\dot{H}}{3H^2}.
\end{equation}
For an accelerated universe we have $\omega<-1/3$. When
$-1<\omega<-1/3$, the universe is in quintessence phase and when
$\omega<-1$, the universe is in phantom phase. In the following,
we assume that the dark matter and dark energy components can
interact through the following source term:
\begin{equation}\label{n1}
 Q=\lambda_m H\rho_m+\lambda_dH\rho_d,
\end{equation}
where $\lambda_m$ and $\lambda_d$ are two real constants. For
special choices such as $\lambda_m=0$, $\lambda_d=0$ or
$\lambda_d=\lambda_m$, eq.(\ref{n1}) reduces to the interaction
terms which have been considered before \cite{inter}. The other
forms of interaction terms, not necessarily suitable for our
purpose, have been also considered in the literature \cite{Qterm}.

Because of the interaction term, we have not the conservation of
partial stress-energy tensors of matter and dark-energy:
$T^{\,\,\mu\nu}_{(m)\,\,\, ;\nu}=-T^{\,\,\mu\nu}_{(d)\,\,\,
;\nu}\neq 0$. In fact, the projection of this nonconservation
equation along the velocity of the whole (comoving) fluid $U_\nu$
(which was taken to be the same as the velocities of each of the
fluid components) is \cite{LP}
\begin{equation}\label{ref1}
 U_\nu T^{\,\,\mu\nu}_{(m)\,\,\, ;\mu}
 =-U_\nu T^{\,\,\mu\nu}_{(d)\,\,\, ;\mu}=
 -Q.
\end{equation}
Note that the coupling (\ref{n1}) can be written as a scalar as
follows
\begin{equation}
 Q=\frac{1}{3}U_\mu U_\nu(\lambda_mT^{\,\,\mu\nu}_{(m)}+\lambda_d
T^{\,\,\mu\nu}_{(d)})U^\alpha_{;\alpha}.
\end{equation}
For FLRW metric, the equation (\ref{ref1}) reduces to:
\begin{eqnarray}\label{4}
\dot{\rho_d}+3H(\rho_d+P_d)&=&-Q,\nonumber \\
\dot{\rho_m}+3H\rho_m&=&Q.
\end{eqnarray}
Using eq.(\ref{1}), eq.(\ref{4}) can be written as
\begin{eqnarray}\label{5}
\dot{\rho_d}+(3+\lambda_d-\lambda_m)H\rho_d
+3HP_d&=&-\frac{3}{8\pi}\lambda_mH^3,
\nonumber \\
\dot{\rho_m}+(3+\lambda_d-\lambda_m)H\rho_m&=&\frac{3}{8\pi}\lambda_dH^3.
\end{eqnarray}
Using eq.(\ref{5}), the evolution equation of the ratio of energy
densities of dark matter and dark energy, denoted by
$r=\rho_m/\rho_d$, reads
\begin{equation}\label{6}
\dot{r}=r(r+1)\left(3\omega+\lambda_m+\frac{\lambda_d}{r}\right)H.
\end{equation}
From
\begin{equation}\label{7}
\Omega_d=\frac{1}{1+r},
\end{equation}
eq.(\ref{6}) then results in
\begin{equation}\label{8}
\omega=-\frac{1}{3H}\frac{\dot{\Omega_d}}{1-\Omega_d}-\frac{\lambda_d\Omega_d}{3(1-\Omega_d)}
-\frac{\lambda_m}{3}.
\end{equation}

In the vicinity of transition time from quintessence to phantom
era, $\omega>-1$ goes to $\omega<-1$, so $\dot{H}$ must change
sign from $\dot{H}<0$ to $\dot{H}>0$. At transition time we have
$\dot{H}=0$ and $\omega=-1$. The dark energy equation of state
parameter $\omega_d$ is defined through $P_d=\omega_d\rho_d$.
Therefore $\omega\rho=\omega_d\rho_d$ or
$\Omega_d\omega_d=\omega$. Using
\begin{eqnarray}\label{9}
\rho_m&=&\frac{3\omega_dH^2+2\dot{H}+3H^2}{8\pi\omega_d}\nonumber
\\
\rho_d&=&-\frac{2\dot{H}+3H^2}{8\pi\omega_d},
\end{eqnarray}
and eq.(\ref{5}), one can obtain the following equation for the
Hubble expansion
\begin{eqnarray}\label{10}
&&\ddot{H}+(6+\lambda_d-\lambda_m+3\omega_d)H\dot{H}+\frac{3}{2}\left[
(3- \lambda_m)\omega_d+3+\lambda_d-\lambda_m\right] H^3\nonumber
\\
&=&\frac{\dot{\omega_d}}{\omega_d}\left(
\dot{H}+\frac{3H^2}{2}\right).
\end{eqnarray}
For a constant $\omega_d$, we arrive at the result of
\cite{barrow}. At $\dot{H}=0$ we obtain
\begin{equation}\label{11}
\ddot{H}=-\frac{3}{2}\left[(3-
\lambda_m)\omega_d+3+\lambda_d-\lambda_m\right]H^3.
\end{equation}
Note that $H>0$, therefore for a constant $\omega_d$, the sign of
$\ddot{H}$ does not change. This shows that in constant-$\omega_d$
approximation, the system can cross the $\omega=-1$ line only
once. This is because the transition from quintessence to phantom
phase needs positive $\ddot{H}$ (at transition time), while the
vice versa needs negative $\ddot{H}$. In the following, we
consider $\omega_d$ as a function of time.

At transition time, we obtain from eq.(\ref{8})
\begin{equation}\label{12}
-3\dot{\omega}=(3-\lambda_m+\lambda_d)\frac{\dot{\Omega_d}}{1-\Omega_d}
+\frac{\ddot{\Omega_d}}{H(1-\Omega_d)},
\end{equation}
which results in
\begin{equation}\label{13}
(3-\lambda_m+\lambda_d)\dot{\Omega_d}
+\frac{\ddot{\Omega_d}}{H}\geq0.
\end{equation}
Insertion of
\begin{eqnarray}\label{14}
\dot{\Omega_d}&=&\frac{8\pi}{3}\frac{\dot{\rho_d}}{H^2},\nonumber \\
\ddot{\Omega_d}&=&\frac{8\pi}{3H^2}\left(\ddot{\rho_d}-2\frac{\ddot{H}}{H}\rho_d\right),
\end{eqnarray}
into eq.(\ref{13}), leads to
\begin{equation}\label{15}
(3-\lambda_m+\lambda_d)H\dot{\rho_d}+\ddot{\rho_d}-\frac{2\ddot{H}}{H}\rho_d\geq
0,
\end{equation}
at transition time.

We assume that, in the neighborhood of transition time, the Hubble
parameter is a differentiable function of time. The Taylor
expansion of $H$ at transition time, which we take as $t=0$, can
be written as \cite{moh}
\begin{equation}\label{16}
H=h_0+h_1t^\a+O(t^{\a+1}),\,\, \a\geq 2, h_1\neq 0.
\end{equation}
 $h_0=H(t=0)$, $\a$ is the order of the first nonzero
derivative of the Hubble parameter at transition time, and
$h_1=(1/\a!)d^\a H/dt^\a |_{t=0}$. The transition from
quintessence to phantom phase occurs if $\a$ is an even positive
integer and $h_1>0$. We also consider the following expansions for
$\Omega_d$, $\rho_m$, and $\rho_d$ at $t=0$
\begin{eqnarray}\label{17}
\Omega_d&=&u_0+u_1t^\b+O(t^{\b+1}),\nonumber \\
\rho_m&=&\rho_{m0}+\rho_{m1}t^\c+O(t^{\c+1}),\nonumber \\
\rho_d&=&\rho_{d0}+\rho_{d1}t^\d+O(t^{\d+1}),
\end{eqnarray}
respectively. $\b$, $\c$, and $\d$ are the orders of the first
nonzero derivatives of $\Omega_d$, $\rho_m$, and $\rho_d$ at
$t=0$, respectively. Note that the above expansions are valid
until $t<<h_0^{-1}$, which is completely reasonable since
$h_0^{-1}$ is of order of age of our universe.

To obtain the relation between the parameters $\a,\,\b,\,\c,$ and
$\d$, we proceed as follows. For $\b\neq 1$, if we expand both
sides of eq.(\ref{8}) at $t=0$, the first resulting term of the
right hand side, with nonvanishing power of $t$, is $t^{\b-1}$
while the left hand side (after $t^0$) begins with $t^{\a-1}$. So
if $\b\neq 1$, we must have $\a=\b$. In this case,
$(3/8\pi)H^2\Omega_d=\rho_d$ results $\d=\b(=\a)$. For $\b=1$,
this equation results in $\d=\b(=1)$. Therefore always $\d=\b$.
From eq.(\ref{9}) it is clear that, for a constant $\omega_d$, we
must have $\b=\a-1$ which leads to $\b=1$ and $\a=2$.

In the case $\b\neq 1$, comparing the coefficients of $t^0$-terms
of eq.(\ref{8}) gives
\begin{equation}\label{100}
u_0=\frac{3-\lambda_m}{\lambda_d-\lambda_m+3},
\end{equation}
and equating the coefficients of $t^{\a-1}$-terms results in
\begin{equation}\label{101}
h_1=\frac{h_0u_1}{2(1-u_0)}.
\end{equation}
This relation shows that the transition is possible only if
$u_1>0$.

In the case $\b=1$, the same procedure leads to
\begin{equation}\label{103}
u_1=(\lambda_m-\lambda_d-3)h_0u_0+(3-\lambda_m)h_0,
\end{equation}
\begin{equation}\label{104}
(\lambda_d-\lambda_m+3)h_0u_{k}+(k+1)u_{k+1}=0, \,\,\, 1<k\leq
\a-2,
\end{equation}
and
\begin{equation}\label{105}
h_1=\frac{(\lambda_d-\lambda_m+3)h_0^2u_{\a-1}+\a
h_0u_\a}{2\a(1-u_0)}.
\end{equation}

The Taylor expansion of $r$ at $t=0$ is
\begin{equation}\label{106}
r=r_0+r_1t^\b+O(t^2),
\end{equation}
where $r_0=u_0^{-1}-1$ and $r_1=-u_1/(u_0^2)$.

\section{Interacting phantom dark energy model and coincidence problem}
In this section we assume that the origin of the dark energy is a
phantom scalar field $\phi$. So
\begin{eqnarray}\label{26}
\rho_d&=&-\frac{1}{2}\dot{\phi}^2+V(\phi),\nonumber \\
P_d&=&-\frac{1}{2}\dot{\phi}^2-V(\phi),
\end{eqnarray}
where $V(\phi)>0$ is the phantom potential.  $\omega_d$ is given
by
\begin{equation}\label{27}
\omega_d=\frac{-\frac{1}{2}\dot{\phi}^2-V(\phi)}{-\frac{1}{2}\dot{\phi}^2+V(\phi)},
\end{equation}
therefore $\omega_d<-1$. For $\Omega_d\omega_d<-1$, the universe
is in the phantom and for $\Omega_d\omega_d>-1$ it is in the
quintessence phase.

The field equation of $\phi$ is
\begin{equation}\label{28}
\dot{\phi}(\ddot{\phi}+3H\dot{\phi}-\frac{dV}{d\phi})=Q.
\end{equation}
This can be derived by putting eq.(\ref{26}) back into
eq.(\ref{4}). From eq.(\ref{26}) we obtain
\begin{eqnarray}\label{29}
\dot{\phi}^2&=&-(1+\omega_d)\rho_d \nonumber \\
2V(\phi)&=&(1-\omega_d)\rho_d.
\end{eqnarray}
The second equation of (\ref{29}) can be written as
\begin{equation}\label{30}
\dot{\phi}=\frac{dV^{-1}(y)}{dy}\dot{y},
\end{equation}
where $y=(1-\omega_d)\rho_d/2$, and $V^{-1}$ is the inverse
function of $V$. Eq.(\ref{30}) and the first equation of
(\ref{29}) then lead to
\begin{equation}\label{31}
(1+\omega_d)\rho_d=-\left(\frac{dV^{-1}(y)}{dy}\right)^2\dot{y}^2,
\end{equation}
which after some calculation can be rewritten as
\begin{equation}\label{32}
\left[\dot{\Omega_d}-\dot{\omega}-3H(1+\omega)(\Omega_d-\omega)\right]^2\left(
\frac{dV^{-1}(y)}{dy}\right)^2=-\frac{4}{\rho}(\Omega_d +\omega).
\end{equation}
This equation together with our previous results may be served to
find some necessary conditions for $\omega=-1$ crossing in
interacting phantom dark energy models, including the domain to
which $u_0$ belongs. We will try to obtain a relation between the
coincidence problem and the behavior of the system at transition
time. For example, for the case $\b=1$, if one obtains $h_1$ as a
polynomial of $u_0$, then restricting $h_1$ to positive values,
which is necessary for transition, will restrict the value of
$u_0$ to a subset of $(0,1)$. By choosing the appropriate
parameters, then it becomes possible to prevent $r$ to be $0$ or
very large. For $\b\neq 1$ cases, eq.(\ref{100}) determines the
value of $r$ at transition time, which again can be chosen to be
$O(1)$. In this way the occurrence of $\omega=-1$ crossing {\it
and} the alleviation of the coincidence problem can be achieved
{\it simultaneously}.

In the following, we will show these points via some specific
examples. In these examples we restrict ourselves to the case
$\a=2$.
\subsection{ Phantom field with square power law potential}

For $V(\phi)=(1/2)m^2\phi^2$,  eq.(\ref{32}) becomes
\begin{equation}\label{33}
2m\sqrt{\omega^2-\Omega_d^2}=\pm\left[\dot{\Omega_d}-\dot{\omega}-3H(1+\omega)(\Omega_d-\omega)\right].
\end{equation}
In the following, we adopt that, in the quintessence phase and
near the transition time, $\dot{\Omega_d}>0$ or equivalently
$r_1<0$ \cite{abd}. Therefore
\begin{equation}\label{34}
2m\sqrt{\omega^2-\Omega_d^2}=\dot{\Omega_d}-\dot{\omega}-3H(1+\omega)(\Omega_d-\omega)
\end{equation}
can be used in the neighborhood of transition time. Taking $\b=1$
(the case $\b\neq 1$ will be discussed later), the expansion of
eq.(\ref{34}) at $t=0$ then results in
\begin{eqnarray}\label{35}
&&2m\sqrt{1-u_0^2}-\frac{2m \left( -4h_1+3u_0 u_1h_0^2 \right)
}{3\sqrt{1-u_0^2h_0^2}}t+O \left( t^2 \right)\nonumber \\
&=& u_1+\frac {4h_1}{3h_0^2}+ \left( 2u_2+4\frac
{h_2}{h_0^2}+4\frac {h_1 \left( 1+u_0
 \right)}{ h_0}\right) t+O \left( t^2 \right).
\end{eqnarray}
As a result we arrive at
\begin{equation}\label{36}
2m\sqrt{1-u_0^2}=u_1+\frac{4h_1}{3h_0^2}
\end{equation}
The necessity of quintessence to phantom phase transition, i.e.
$h_1>0$, then results in
\begin{equation}\label{37}
u_1>2m\sqrt{1-u_0^2}.
\end{equation}
Using eq.(\ref{103}), we can write the above inequality in terms
of $u_0$:
\begin{equation}\label{38}
au_0+b<c\sqrt{1-u_0^2}.
\end{equation}
We have defined $a=\lambda_m-\lambda_d-3$, $b=3-\lambda_m$, and
$c=2m/h_0$.

To study the solutions of eq.(\ref{38}), we consider two
situations. The first possibility is:
\begin{equation}\label{39}
au_0+b\leq 0,
\end{equation}
which leads to $r_1\geq 0$. This conflicts with the assumption
$r_1 < 0$, or equivalently $\dot{\Omega_d}> 0$ at transition time,
and therefore is not acceptable. The second possibility is
$au_0+b> 0$ which leads to
\begin{equation}\label{40}
\mathcal{P}(u_0):=(a^2+c^2)u_0^2+2abu_0+b^2-c^2<0.
\end{equation}
If $a^2-b^2+c^2<0$, $\mathcal{P}$ has no real roots and its sign
does not change. But $\mathcal{P}(1)>0$, therefore eq.(\ref{40})
is not satisfied in this case. For $a^2-b^2+c^2>0$, $\mathcal{P}$
has two roots which we denote by $u_{R1}$ and $u_{R2}$.
Eq.(\ref{40}) is satisfied if the value of $\Omega_d$ at
transition time is restricted to the intersection of the intervals
$(u_{R1},u_{R2})$ and $(0,1)$
\begin{equation}\label{41}
u_0\in (0,1)\bigcap (u_{R1},u_{R2}).
\end{equation}
So if $(u_{R1},u_{R2})\subset (0,1)$, by choosing the appropriate
parameters $a,b,$ and $c$, we can obtain the desired order of
magnitude: $\sim O(1)$ for $r_0=1/{u_0}-1$. The Sturm sequences at
$0$ and $1$ are:
\begin{equation}\label{n2}
S(0)=\left[b^2-c^2,\, 2ab,\, \frac{c^2(a^2-b^2+c^2)}{
a^2+c^2}\right],
\end{equation}
 and
\begin{equation}\label{n3}
S(1)=\left[(a+b)^2,\, 2(a^2+c^2+ab),\, \frac{c^2(a^2-b^2+c^2)}{
a^2+c^2}\right].
\end{equation}
Using Sturm theorem, one can show that for
\begin{eqnarray}\label{42}
&&a^2-b^2+c^2>0,\nonumber \\
&&a^2+c^2+ba>0,\nonumber \\
&&b^2-c^2>0,\nonumber \\
&&ab<0,
\end{eqnarray}
the two roots of $\mathcal{P}$ belong to $(0,1)$. In this way we
have
\begin{equation}\label{43}
\frac{-ab-c\sqrt{a^2+c^2-b^2}}{a^2+c^2}<\Omega_d<\frac{-ab+c\sqrt{a^2+c^2-b^2}}{a^2+c^2},
\end{equation}
at transition time. As an example, consider the case
$\lambda_m=1$, $\lambda_d=2$ and $c=1$. In this case
$u_{R1}=0.258$ and $u_{R2}=0.682$, therefore $0.46<r<2.8$ at
transition time. Note that $(u_{R1},u_{R2})$ may be more tightened
by choosing appropriate $a$, $b$ and $c$, e.g. for $c=1$, $a=7$,
and $b=-5$, which correspond to $\lambda_m=8$ and $\lambda_d=-2$,
we have $u_{R1}=0.6$ and $u_{R2}=0.8$, therefore $0.6<u_0<0.8$ in
agreement with the expected value $u_0=0.7$ and $r_0=3/7$.

In $\b\neq 1$ cases, we have $\b=\a$. For $\a=2$, following the
same method ending to eq.(\ref{35}), we expand both sides of
eq.(\ref{34}). It is obtained, up to the order $O(t)$,

\begin{equation}\label{108}
m\sqrt{1-u_0^2}=\frac{2h_1}{3h_0^2}
\end{equation}
which results in $h_1>0$. $u_0$ and $u_1$ are given by
eqs.(\ref{100}) and (\ref{101}), respectively. Higher orders of
$t$ in the Taylor expansion of eq.(\ref{34}) determine the other
coefficients of the Taylor expansion of $H$ an $\Omega_d$. Since
$h_1>0$ induces no additional constraint on $u_0$, the appropriate
parameters $a,\, b$, and $c$ can lead to the desired values for
$r_0$. For $\a>2$, the aforementioned expansion leads to
$m\sqrt{1-u_0^2}=0$ which is ruled out by  the assumption that
$u_0\neq 1$. Therefore in $\b\neq 1$, the choice $\a=2$ is the
only eligible one.

Eq.(\ref{34}) with $\dot{\Omega}_d<0$ and eq.(\ref{33}) with minus
sign can be also investigated by the same method. In brief, it is
shown that, in interacting phantom model with
$V(\phi)=(1/2)m^2\phi^2$ phantom potential and the interaction
$Q$-term of eq.(\ref{n1}), it is possible to choose the parameters
such that {\it both} the $\omega=-1$ crossing and $r_0=O(1)$
occur. In special case which leads to eq.(\ref{108}), one can tune
the parameters such that $r_0$ has no choice but the desired value
3/7.

\subsection{Exponential potential}
Consider the following  potential\label{44}
\begin{equation}
V=v_0\exp(\lambda \phi), \,\, \lambda>0,\,\, v_0>0.
\end{equation}
Eq.(\ref{32}) then results in
\begin{equation}\label{45}
H\tilde{\lambda}(\omega^2-\Omega_d^2)^{\frac{1}{2}}(\Omega_d-\omega)^{\frac{1}{2}}
=\pm\left[\dot{\Omega_d}-\dot{\omega}-3H(1+\omega)(\Omega_d-\omega)\right],
\end{equation}
where $\tilde{\lambda}=\sqrt{3/(8\pi)}\lambda$. As the previous
example, we consider the upper sign of eq.(\ref{45}) which is a
result of the assumptions $\dot{\Omega_d}>0$ and $\dot{\omega}<0$
in the vicinity of transition time.

By Taylor expansion of the both sides of eq.(\ref{45}) at
transition time, we obtain the following equation for $\a=2$ and
$\b=1$
\begin{equation}\label{46}
\frac{4h_1}{3h_0^3}=-au_0-b+\tilde{\lambda}(1-u_0^2)^{\frac{1}{2}}(1+u_0)^{\frac{1}{2}}.
\end{equation}
$a$ and $b$ are defined by the same relations as for the first
example. $h_1>0$ then results in
\begin{equation}\label{47}
au_0+b<\tilde{\lambda}(1-u_0^2)^{\frac{1}{2}}(1+u_0)^{\frac{1}{2}}.
\end{equation}
Eq.(\ref{47}) is satisfied in two cases: (i) $au_0+b<0$, which
makes $r_1$ negative, or $\dot{\Omega_d}<0$, and is not
acceptable. (ii) $au_0+b>0$. In this case we must have
\begin{equation}\label{48}
u_0^3+(A^2+1)u_0^2+(2AB-1)u_0+B^2-1<0,
\end{equation}
where $A=a/(\tilde{\lambda}^2)$ and $B=b/(\tilde{\lambda}^2)$. We
also assume $B^2>1$. Eq.(\ref{48}) is satisfied only if the
polynomial
 \begin{equation}\label{49}
\mathcal{Q}(u_0):=u_0^3+(A^2+1)u_0^2+(2AB-1)u_0+B^2-1
\end{equation}
has real roots. Following Descartes rule,  $B^2-1>0$ and $2AB-1<0$
are necessary conditions for $\mathcal{Q}(u_0)$ to have two real
positive roots. The domain to which $u_0$ in eq.(\ref{48}) belongs
is the intersection of $(0,1)$ and $(u_{R1},u_{R2})$, where
$u_{R1}$ and $u_{R2}$ are the roots of $\mathcal{Q}(u_0)$. So by
appropriate choosing of $A$ and $B$, one can restrict $u_0$ to the
domain allowed by astrophysical data. To do so, we construct the
Sturm sequence corresponding to the cubic polynomial
$\mathcal{Q}(u_0)$ at 0 and 1. They are
\begin{eqnarray}\label{51}
S(0)=\Big[&& B^2-1,\, 2AB-1,\, \frac{1}{9}(2AB-1)(A^2+1)+(1-B^2),
\nonumber\\ && \frac{9}{4}\,\frac{D}{( A^4+2A^2-6AB+4)^2}\Big],
\end{eqnarray}
and
\begin{eqnarray}\label{52}
S(1)=\Big[ &&A^2+2AB+B^2,\, 2A^2+2AB+4,\,
\frac{1}{9}(2A^4+2A^3B+3A^2
\nonumber\\&&-10AB-9B^2+16),\,\frac{9}{4}\,\frac{D}{(
A^4+2A^2-6AB+4)^2}\Big].
\end{eqnarray}
In above equations, $D>0$ is discriminant of the polynomial
$\mathcal{Q}(u_0)$. By implying Sturm theorem, it can be verified
that, in order to have two real roots in  interval $(0,1)$, the
parameters $A$ and $B$ must satisfy, besides the previous
mentioned conditions $B^2-1>0$ and $2AB-1<0$, the following
inequalities
\begin{eqnarray}\label{53}
&&A^2+AB+2>0,\nonumber \\
&&2A^4+2A^3B+3A^2-10AB-9B^2+16>0.
\end{eqnarray}
For example, for $\tilde{\lambda}=1$, $\lambda_m=1$, and
$\lambda_d=2$, we obtain $0.23<u_0<0.73$ which is in agreement
with $u_0\sim 0.7$ obtained from astrophysical data.

Now we consider $\b\neq 1$. For $\a=2$, the Taylor expansion of
the both sides of eq.(\ref{45}), with upper sign, result in
\begin{eqnarray}
&&\frac{4}{3}\,{\frac
{h_{{1}}}{{h_{{0}}}^{2}}}-\tilde{\lambda}\,\sqrt
{1-{u_{{0}}}^{2}} \sqrt {1+u_{{0}}}+ \nonumber \\
&&\left( 4\,{\frac {h_{{2}}}{{h_{{0}}}^{2}}}-2\,u_{{2 }}+4\,{\frac
{h_{{1}} \left(1+u_{{0}}\right) }{h_{{0}}}}-{2\over 3}\,{ \frac
{\tilde{\lambda}h_1\sqrt {1-{u_{{0}}}^{2}}}{h_0^2\sqrt
{1+u_{{0}}}}}-{4\over 3}\,{\frac {\tilde{\lambda}h_{{1}}\sqrt
{1+u_{{0}}}}{h_0^2\sqrt
{1-{u_{ {0}}}^{2}}}} \right) t+\nonumber \\
&& O \left( {t}^{2} \right)=0.
\end{eqnarray}
Therefore
\begin{equation}\label{109}
\frac{4h_1}{3h_0^3}=\tilde{\lambda}(1-u_0^2)^{\frac{1}{2}}(1+u_0)^{\frac{1}{2}}
\end{equation}
which implies $h_1>0$. By suitable choosing of $\lambda_m$ and
$\lambda_d$, one can obtain the appropriate value for $r$ at
transition time. For $\a>2$, the aforementioned expansion leads to
$\tilde{\lambda}(1-u_0^2)^{\frac{1}{2}}(1+u_0)^{\frac{1}{2}}=0$
which is ruled out by  the assumption that $u_0\neq 1$. Therefore
$\a=2$ is the only allowed case for $\b\neq 1$.

\section{Conclusion}
In this paper, by considering the energy exchange between cold
dark matter and dark energy (see eq.(\ref{n1})), we study the
possibility of simultaneous occurrence of two phenomena, the
coincidence problem, and $\omega =-1$ crossing, from $\omega
>-1$ to $\omega <-1$. We consider the physical quantities near
the transition time $t=0$, through eqs.(\ref{16}) and (\ref{17}).
The transition occurs for positive $h_1$ and even $\alpha$, the
parameters which has been introduced in eq.(\ref{16}).

The equation of state parameter $\omega$ is expressed by
eq.(\ref{8}) and the potential of phantom field, as a candidate of
dark energy, enters in eq.(\ref{32}). We studied the perturbative
solutions of these equations, near $t=0$, for two specific
potentials, i.e. the quadratic and exponential potential. It is
shown that always $\theta=\beta$, and for $\beta\neq 1$,
$\alpha=\beta$ (see eq.(\ref{17}) and its subsequent discussion).
For $\alpha=2$, as a first acceptable solution for $\omega=-1$
crossing, it is shown that, in both examples, it is possible to
choose the parameters such that, besides the satisfaction of
dynamical equations, the occurrence of $\omega
>-1$ to $\omega <-1$ transition allows the ratio
$r_0=(\rho_m/\rho_d)_{t=0}$ to be around the desired value 3/7.
This proves the possibility of solving these two problems in a
unique framework.

For $\alpha>2$ and $\beta=1$, it can also be shown that it is
possible to choose the parameters such that the above mentioned
properties are achieved.

 {\bf Acknowledgement:} We would like to thank M. Honardoost
for useful discussion. This work was partially supported by the
"center of excellence in structure of matter" of the Department of
Physics.

\end{document}